\documentclass[doublecol]{epl2}

\title{Gap state charge induced spin-dependent negative differential resistance in tunnel junctions}
\shorttitle{Gap state charge induced spin-dependent negative differential resistance in tunnel junctions}

\author{Jun Jiang\inst{1,2} \and X.-G. Zhang\inst{2} \and X. F. Han\inst{1}\thanks{E-mail:\email{xfhan@iphy.ac.cn}}}
\shortauthor{Jun Jiang \etal}

\institute{
  \inst{1} Beijing National Laboratory for Condensed Matter Physics, Institute of Physics, Chinese Academy of Sciences, Beijing 100190, China\\
  \inst{2} Department of Physics and the Quantum Theory Project,University of Florida, Gainesville, Florida 32611, USA
}

\abstract{
We propose and demonstrate through first-principles calculation a new spin-dependent negative differential resistance (NDR) mechanism in magnetic tunnel junctions (MTJ) with cubic cation disordered crystals (CCDC) AlO$_x$ or Mg$_{1-x}$Al$_x$O as barrier materials. The CCDC is a class of insulators whose band gap can be changed by cation doping. The gap becomes arched in an ultrathin layer due to the space charge formed from metal-induced gap states. With an appropriate combination of an arched gap and a bias voltage, NDR can be produced in either spin channel. This mechanism is applicable to 2D and 3D ultrathin junctions with a sufficiently small band gap that forms a large space charge. It provides a new way of controlling the spin-dependent transport in spintronic devices by an electric field. A generalized Simmons formula for tunneling current through junction with an arched gap is derived to show the general conditions under which ultrathin junctions may exhibit NDR.}

\begin{document}

\maketitle

\section{Introduction}
Negative differential resistance (NDR) is one of the most important nonlinear electronic properties,  as evident from the broad range of recent studies covering graphene-based ballistic field-effect transistor \cite{dragoman2014negative}, p-GaN/Mg-doped Al$_{0.15}$Ga$_{0.85}$N/n-GaN hetero-junction \cite{zhang2014negative} and single-layer MoS$_2$ with a tunable gate \cite{sangwan2015gate}. A vibronic mechanism for NDR was predicted in molecular junctions \cite{PhysRevB.83.115414}. NDR is also predicted in salicylideneaniline molecular sandwiched devices \cite{xie2015negative}, quantum dots-based organic light-emitting diodes\cite{yang2014negative} and planar graphene nanoribbon break junctions \cite{nguyen2015negative}.
Devices for generating NDR mostly use semiconductor diodes \cite{Esaki,Gunn1964,esaki1966new} and tunnel junctions \cite{Tsu1973,Chang1974}, and are used in power amplifiers, oscillators, and in switching circuits.
Popular NDR mechanisms mostly rely on bulk effects or require large space charge regions that present challenges for scaling the device down to the nanoscale.
Additional challenges are to perform similar functions in 2D materials \cite{BandgapGraphene,GrapheneTransistors,GrapheneTransport,BiasedBilayerGraphene,Kim} and for spin currents in spintronics. Spin-dependent NDR was predicted for double barrier junctions
\cite{PhysRevB.73.033409} and for magnetic quantum well layers \cite{PhysRevLett.94.207210}, but the resonant conditions required for such effect are difficult to achieve. Spin
blockade in quantum dots can also lead to NDR \cite{CiorgaAPL2002}.

In this paper, we show that ultrathin junctions with barriers about 1 nm in thickness can be exploited to produce NDR through a simple yet unexplored mechanism. When an insulator material has a small band gap, the metallic electronic states from electrodes can penetrate the barrier layer. This causes a space charge due to the metal-induced gap state (MIGS) \cite{PhysRevB.32.6968} to build up in the barrier layer, which in turn produces an internal electric field. The resulting shifts of the effective band edges cause the band gap to become ``arched'' across the thickness of the barrier. Such an arched band gap was experimentally observed in Fe/GaAs/Fe \cite{PhysRevLett.114.146804}. The arch makes the Fermi energy stay close to (or may even enter) either the conduction band or the valence band of the barrier at the interfaces with the electrodes. The effective barrier thickness at zero bias is therefore smaller than the nominal thickness. When a bias voltage is applied, at each interface the band
edge is pinned at the corresponding electrochemical potential. As the bias is increased, while the effective barrier thickness continuously decreases, the effective barrier height for each tunneling state within the transport energy window varies in a nonlinear manner. With the right combination of the arched band gap and the applied bias, the current may decrease with the bias, leading to a NDR.

\begin{figure}[htbp]
\centering
\includegraphics[width=0.44\textwidth]{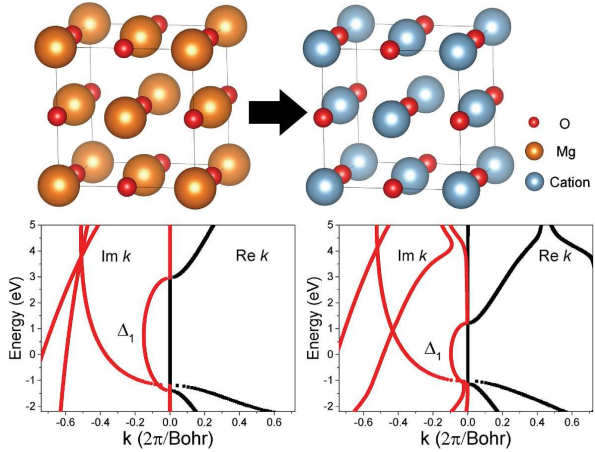}
\caption{\label{structure} Top: Structures of MgO (left) and a CCDC material (right), where the cation sites are randomly filled by the two cations (Mg, Al) for MgAlO, (Al, vacancy) for AlO, or (Mg, Zn) for MgZnO. Bottom: Complex band structures of MgO (left) and CCDC-alumina (right) along the (001) direction.}
\end{figure}

The traditional MgO barrier does not show any significant space charge region from either theoretical calculation \cite{PhysRevB.63.054416} or experiments \cite{Parkin2004,Yuasa2004}. The 9 eV MgO band gap is too large to allow a charge density from MIGS. The gap can be
 reduced by Zn or Al doping \cite{Li2014srep, ZhangJiaAPL}. Another material with a small and tunable band gap is $\gamma$-alumina, which
 can be considered as based on a spinel structure with substitutional vacancies replacing Al atoms. The spinel like Al$_2$O$_3$ (100) film has been made experimentally and is stable under atmosphere\cite{spinelAlO}.
 Cubic spinel materials are shown to have $\Delta_1$ spin filter feature similar to MgO \cite{ZhangJiaAPL}. However the unit cell of an ordered spinel with two types of cations is double that of MgO,
which causes the $\Delta_1$ band to be folded such that minority spin channel has $\Delta_1$ electrons at the fermi energy, diminishing the spin filtering effect \cite{PhysRevB.86.184401}.  Cation disordered crystals, in which the cation sites are substitutionally disordered, maintain the cubic symmetry in the
 smaller unit cell thus avoiding the band folding problem.\cite{PhysRevB.86.184401}
 Therefore, we will consider materials made from random substitutions of Mg atoms in MgO by Zn or Al atoms, and of Al atoms in AlO (in the structure of MgO) by vacancies, as illustrated in Fig. \ref{structure}.
For the latter to have the stoichiometry of Al$_2$O$_3$, one third of the Al atoms are randomly replaced by vacancies.
 For brevity, in the rest of the paper we will use generic names AlO, MgAlO, and MgZnO to refer to the respective CCDC materials.

\section{Models and method}

The first-principles band structure and transport calculations use the Layer-Korringa-Kohn-Rostoker (LKKR) code \cite{PhysRevB.59.5470} employing the density functional theory (DFT) with the local spin density approximation \cite{vosko1980accurate} and the Landauer transport formula \cite{buttiker1985generalized}.  Substitutional disorder due to vacancies in AlO as well as Al and Zn doping in MgAlO and MgZnO is treated with the coherent-potential-approximation (CPA) \cite{PhysRevB.5.2382}. The lattice parameter of the bcc Fe is set to the experimental bulk value of 2.866 {\AA}. The distances between the interface Fe layer and the first CCDC layer and between the CCDC layers are set to the same values as in the MgO based MTJ, at 2.163 {\AA} and 2.210 {\AA}, respectively. The ratio between the radii of the O atom and of the cation atoms and the vacancy under the atomic sphere approximation (ASA) is set to be the same as that between O and Mg in the MgO structure in earlier works \cite{PhysRevB.63.054416,Li2014srep}.

The space charge in the barrier layer is calculated using the MIGS model, in which we assume that the space charge density is formed from the exponentially decaying electron wave functions within the band gap, in the form $\psi(E,z)\approx A e^{-\int\widetilde{\kappa}dz}$ at energy $E$,
where $\widetilde{\kappa}=\widetilde{\kappa}[E-eU(z)]$ is the local decay rate of the wave function determined from the first-principles complex band structure
of the material as described below, $U(z)$ is the local electrostatic
potential, and $z$ is the direction perpendicular to the junction plane. Using these wave functions we find,
\begin{eqnarray}
\rho(z)&=&[1+\cos(2k_rz+\delta)]\left[
\int_{E_v+eU(z)}^{\mu_L}dE \rho_Le^{-2\int_0^z\widetilde{\kappa}dz'}\right.\nonumber\\
&+&\left.\int_{E_v+eU(z)}^{\mu_R}dE\rho_Re^{-2\int_d^z\widetilde{\kappa}dz'}\right],
\label{rhoz}
\end{eqnarray}
where $E_v$ is the valence band edge, $d$ is the
thickness of the barrier, and $\rho_{L(R)}$
provide the boundary conditions at the two interfaces for the charge.
The oscillatory cosine term is added to account for charge oscillations in the barrier layer that appear in the first-principles
calculations for some of the junctions and is not intrinsic to the MIGS model. The parameters $k_r$ and $\delta$ are fitted
from first-principles results.

For each atomic layer in the barrier, the $\Delta_1$ complex band at energy $E$ has an imaginary part $\kappa(E)$ along the (001) direction. This is extracted from the first-principles
complex band structure calculation. Within the
WKB approximation, the Schr\"odinger equation can be written as
$d^2\psi(z)/dz^2-\kappa^2\psi(z)=0$,
where $\kappa$ is now also a function of $E-eU(z)$. The differential equation for
the effective local decay rate $\widetilde{\kappa}[E-eU(z)]$ can be derived as,
\begin{equation}
-\frac{d\widetilde{\kappa}}{dz}+\widetilde{\kappa}^2=\kappa^2,
\end{equation}
where $\kappa(E)$ is given by the two-band effective mass model for a homogeneous band gap \cite{PhysRevB.63.054416},
\begin{equation}
\frac{1}{\kappa^2(E)}=\frac{\hbar^2}{2m_c(E_c-E)}+\frac{\hbar^2}{2m_v(E-E_v)},
\end{equation}
with $E_c$, $m_c$, and $m_v$ being the conduction band edge and effective masses of the conduction and valence bands,
respectively.

In addition to the space charge from the MIGS, other contributions to the electrostatic potential come from a constant surface charge $\sigma$ on the electrodes, the electrostatic potential shift $U_{dp}(z)$  due to the dipole moment of each
layer, and an uncompensated net charge due to the deviation from stoichiometry. The last contribution arises from cation doping (or vacancies) in the barrier layer. In experimental samples, such doping may not always reach stoichiometry. When there is a deviation, charge is injected from the electrodes into the barrier layer in order to
pin either the top of the valence band or the bottom of the conduction band at the Fermi energy of the electrodes. We assume a uniformly distributed $\rho_{bg}$ to account for this charge.
The electrostatic potential $U(z)$ is,
\begin{equation}
U(z)=U_0+U_{dp}(z)+\epsilon_0\int_0^z[\rho(z')+\rho_{bg}]|z-z'|dz'+\epsilon_0\sigma z,
\label{Vz}
\end{equation}
with the boundary conditions $U(0)=U_0=\mu_L-E_F$ and $U(d)=\mu_R-E_F$ (which determines $\sigma$) where $E_F$ is
the Fermi energy at zero bias. Eqs. (\ref{rhoz}) through (\ref{Vz}) are solved simultaneously to yield $\rho(z)$ and
$U(z)$ for each junction at each bias voltage. At zero bias $\rho(z)$ and $U(z)$ from the MIGS model are in perfect agreement
with the self-consistent first-principles results for all junctions considered, validating the model.

The sheet conductance is calculated from the transmission probability \cite{PhysRevB.59.5470},
$G(E)=(e^2/hA)\sum_{{\bf k}_{\|}} T({\bf k}_{\|},E)$.
Under a finite bias $eV=\mu_L-\mu_R$ with $\mu_{L(R)}$ as the
electrochemical potential of the left (right) lead, the current-voltage ($I$-$V$) curves are calculated by shifting the electrostatic potential of the two electrodes by $\pm eV/2$ respectively, and that of the barrier region using the result from Eq. (\ref{Vz}) under the bias boundary conditions. Neglecting vertex corrections from impurity scattering, the current density is given by,
\begin{equation}
 J = \frac{e^2}{hA}\sum_{{\bf k}_{\|}}\int  T({\bf k}_{\|},E)[f(E-\mu_L)-f(E-\mu_R)]dE,
  \label{eq2}
\end{equation}
where $A$ is the area of the junction, and a two-dimensional wave vector
${\bf k}_{\|}$ and the energy parameter $E$ are used to describe a Bloch state. This is a non-self-consistent approximation accurate for low to moderate voltages \cite{PhysRevB.69.134406}.

\begin{figure}[t]
  \centering
  \includegraphics[width=0.46\textwidth]{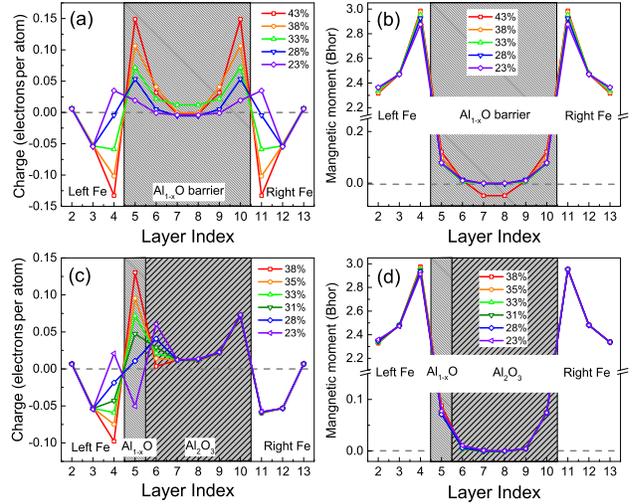}\\
  \caption{The layer-by-layer charge (a), (c) and magnetic moment (b), (d) of the Al$_{1-x}$O based MTJs. In (a) and (b), the cation site vacancies are uniformly distributed and varied. In (c) and (d), only the cation site vacancy concentration on one of the interfacial layers  is varied while all other layers are kept at $1/3$ concentration (that of Al$_2$O$_3$).}
  \label{charge}
\end{figure}

\section{Results and discussion}

\subsection{Metal-induced gap states}

Charge transfer occurs at the interface between a metal and a semiconductor or insulator in order to ensure that the fermi level is matched across the interface. In an ultrathin junction the MIGS are formed from the evanescent electron wave functions that match continuously to the Bloch wave
functions on the metal side. In the MTJs with bcc-Fe(001) as electrode, the main contribution of the barrier charge is from the slowest decaying MIGS with the $\Delta_1$ symmetry.
The decay rates are determined by the complex bands in the gap of the barrier materials, which for MgO and Al$_2$O$_3$ are plotted in Fig. \ref{structure}. These complex bands are calculated using the Kohn-Sham potential of the middle atomic layer of the barrier produced from the self-consistent DFT calculation of the tunnel junction. The
purely imaginary bands are largely unaffected by the disorder, but for Al$_2$O$_3$ all real bands have a small imaginary part of $k_z$. The band gap is found from the real part of $k_z$.

Smaller band gaps tend to allow more MIGS charge leading to more arching. Comparing to the calculated 4.37 eV gap for MgO, the gap for Al$_2$O$_3$ is 2.33 eV. Both numbers are lower than the DFT band gap for bulk materials \cite{PhysRevB.80.144101} because they are calculated using the potentials and the lattice
parameters of the middle layer in the MTJ. Nonetheless, the trend that Al$_2$O$_3$ has a much smaller band gap than MgO is still valid. We also look for good $\Delta_1$ symmetry
filtering for a strong spin-dependent effect. Since both MgO and Al$_2$O$_3$ produce the smallest imaginary part for the $\Delta_1$ bands at the electrode Fermi energy, both have good $\Delta_1$ filtering.
We find that most of the CCDC materials have similar complex band structure to MgO and show good $\Delta_1$ filtering.

In Fig. \ref{charge} we show the layer-by-layer distribution of the MIGS charge and the local magnetic moment in the Al$_{1-x}$O MTJs with different distributions of the cation site vacancies. In Figs. \ref{charge}(a) and (b), the cation sites vacancies are distributed and varied uniformly in the barrier region. When $x\approx 1/3$ in Al$_{1-x}$O (stoichiometry of Al$_2$O$_3$), the barrier layers are significantly charged. When $x$ deviates from 33\% by more than 5\%, the Fermi level moves outside of the gap and the strong screening of the charge at the interface causes the middle part of the barrier to become charge neutral. In a real
junction, the vacancy concentration is often different at the interfaces than in the middle of the barrier region. This situation is considered in Fig. \ref{charge} (c), where the concentration of the cation site vacancy is kept at a constant of $x=1/3$ for all layers except a single interface layer, where $x$ is varied. The interface layer goes from negatively charged to positively charged reflecting the change of the electron affinity. The  magnetic moment has little dependence on the cation sites vacancy concentration, as shown in Fig. \ref{charge} (b) and (d).

\subsection{Arched band gap under finite bias}

\begin{figure}[htbp]
\includegraphics[width=0.46\textwidth]{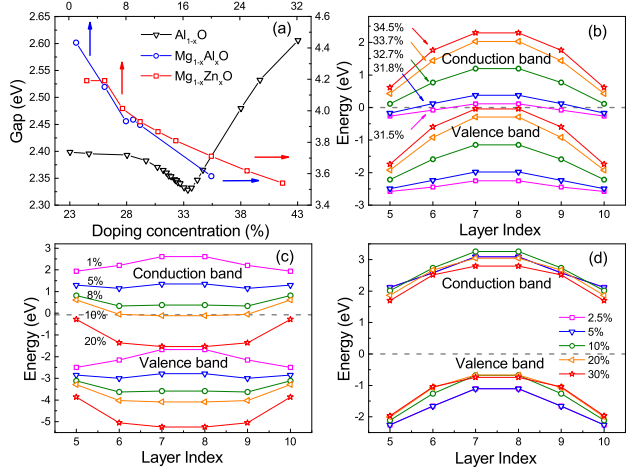}
\caption{\label{gap_band} (a) Band gap of CCDC materials as a function of cation doping. (b-d) Effective conduction and valence band edges for each atomic layer showing arched band gaps for the CCDC materials, (b) Al$_{1-x}$O, (c) Mg$_{1-x}$Al$_{x}$O, and (d) Mg$_{1-x}$Zn$_{x}$O.}
\end{figure}

We first present the band gap values at zero bias as a reference. In Fig. \ref{gap_band}(a) we plot the band gap, extracted from the complex band calculations, as a function of doping (or vacancy) concentration, for the three CCDC materials Al$_{1-x}$O (black), Mg$_{1-x}$Al$_{x}$O (blue), and Mg$_{1-x}$Zn$_{x}$O (red). The band gap is minimum at stoichiometry for AlO, but decreases monotonically for MgAlO and MgZnO.

The arched band gaps at zero bias and under different doping (or vacancy) concentrations in Fe/Al$_{1-x}$O/Fe, Fe/Mg$_{1-x}$Al$_{x}$O/Fe and Fe/Mg$_{1-x}$Zn$_{x}$O/Fe are shown in Fig. \ref{gap_band}(b-d).
In these figures we plot the effective conduction and valence band edges with the $\Delta_1$ symmetry relative to the Fermi energy. The effective band edges are extracted for each atomic layer using its Kohn-Sham potential to form a bulk structure and calculating the corresponding complex band structure. The arched shape reflects a large internal electric field arising from the space charge within the barrier and is consistent with the DOS and charge distribution results presented above. In Fe/Al$_{1-x}$O/Fe and Fe/Mg$_{1-x}$Zn$_{x}$O/Fe MTJs, the barrier region is negatively charged, causing the arch of the band gap edges to bend upward. In Fe/Mg$_{1-x}$Al$_{x}$O/Fe MTJs, the charge in the barrier region can be tuned from negative to positive by changing the Al doping concentration on cation sites. Correspondingly the arch of the band gap edges changes from upward to downward.

\begin{figure}
\centering
\includegraphics[width=0.46\textwidth]{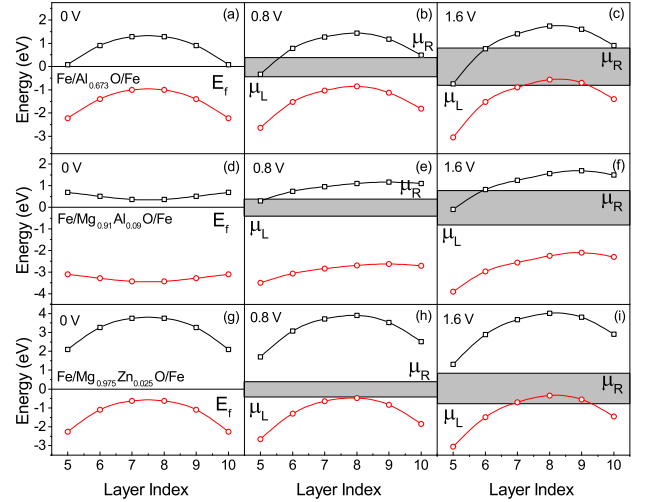}
\caption{\label{MIGS_model} Arched conduction (black) and valence (red) band edges under finite bias  in CCDC MTJs from the MIGS model: (a)-(c) Al$_{0.673}$O; (d)-(f) Mg$_{0.91}$Al$_{0.09}$O; (g)-(i) Mg$_{0.975}$Zn$_{0.025}$O. Shaded boxes indicate the transport energy window bounded by the chemical potentials $\mu_L$ and $\mu_R$ of the two leads.}
\end{figure}

The shifts of the band gap under a finite bias are calculated by the finite-bias MIGS model presented in section II. The charge distribution $\rho(z)$ and the potential shift $U(z)$ are calculated self-consistently making use the complex band structure of the barrier material from first-principles. The model is validated at zero bias by comparing to the charge and potential calculated from first-principles. The band gap edges of Fe/Al$_{0.673}$O/Fe, Fe/Mg$_{0.91}$Al$_{0.09}$O/Fe and Fe/Mg$_{0.975}$Zn$_{0.025}$O/Fe MTJs under biases 0, 0.8, and 1.6 volts are shown in Fig. \ref{MIGS_model}.

\subsection{Spin-dependent negative differential resistance}

\begin{figure}
\centering
\includegraphics[width=0.48\textwidth]{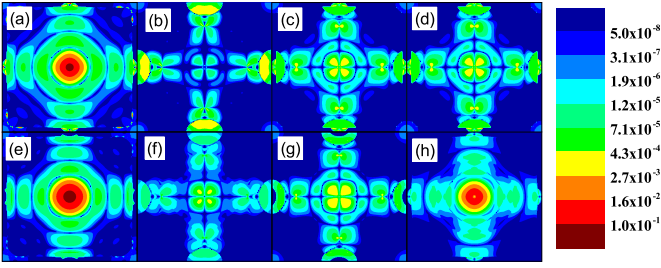}
\caption{\label{TK_bias} Transmission probability in the two-dimensional reciprocal space at the Fermi energy for an Fe/Al$_2$O$_3$/Fe MTJ  under 0 V (top row) and 1 V (bottom row) bias for parallel (a), (b), (e), (f) and anti-parallel (c), (d), (g), (h). (a), (e) majority to majority spin channel; (b), (f) minority to minority spin channel; (c), (g) majority to minority spin channel; (d), (h) minority to majority spin channel.}
\end{figure}

The calculated transmission probability for an Fe/Al$_2$O$_3$/Fe MTJ with (bottom row) and without (top row) bias voltage is plotted in Fig. \ref{TK_bias}. In this figure we show the transmission at the Fermi energy as a function of $k_x,k_y)$ in the two-dimensional reciprocal space for majority to majority (Fig. \ref{TK_bias}(a) and (e)), majority to minority (Fig. \ref{TK_bias}(b) and (f)), minority to majority (Fig. \ref{TK_bias}(c) and (g)) and minority to minority (Fig. \ref{TK_bias}(d) and (h)) spin channels. The strong spin-dependence of the tunneling conductance arises from the dominant majority to majority channel transmission concentrated at the $\bar\Gamma$ point. This is a common behavior of almost all CCDC based MTJs.

\begin{table}
\caption{Zero bias sheet conductance in $1/\mu\Omega$cm$^2$ and TMR of (001) MTJs with barrier materials MgO, Mg$_{0.975}$Zn$_{0.025}$O(6ML), Mg$_{0.91}$Al$_{0.09}$O(6ML),  and CCDC-alumina with 6, 7 and 8 atomic layers in the barrier.}
\begin{tabular}{ccccc}
Barrier (layers) & $G_p^{\uparrow\uparrow}$ & $G_p^{\downarrow\downarrow}$ & $G_{ap}^{\uparrow\downarrow}$ & TMR \\
\hline
MgO (6) & $5.70$ & $0.303 $ & $0.179 $ & 1574\% \\
Mg$_{0.975}$Zn$_{0.025}$O (6) & $5.31$ & $0.572$ & $0.27$ & 991\% \\
Mg$_{0.91}$Al$_{0.09}$O (6) & $152$ & $0.363 $ & $1.65$ & 4526\% \\
CCDC-alumina (6) & $202$ & $2.02$ & $2.38$ & 4184\% \\
CCDC-alumina (7) & $49.1$ & $0.71$ & $0.414$ & 5907\% \\
CCDC-alumina (8) & $15.4$ & $0.302$ & $0.133$ & 5818\% \\
\end{tabular}
\end{table}

Table 1 lists the sheet conductance of the CCDC based MTJs and compares them to the MgO MTJ. All three CCDC barrier materials are effective spin filters. Fe/Mg$_{0.975}$Zn$_{0.025}$O(6ML)/Fe, Fe/Mg$_{0.91}$Al$_{0.09}$O(6ML)/Fe, and Fe/CCDC-alumina (6,7,8 ML)/Fe MTJs have over 900\% TMR due to good $\Delta_1$ filtering. All three materials also have more than 10 times larger sheet conductance than the MgO based MTJ at the same barrier thickness.

\begin{figure}[htbp]
\includegraphics[width=0.46\textwidth]{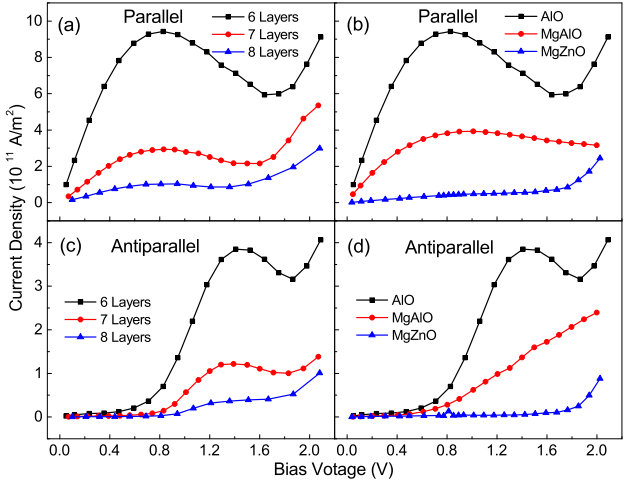}
\caption{\label{IV} (a) $I$-$V$ for parallel moments in Fe/Al$_{1-x}$O/Fe ($x=0.327$) junctions with 6, 7, 8 atomic layers in the barrier; (b) $I$-$V$ curves for parallel moments in Fe/Al$_{1-x}$O/Fe ($x=0.327$), Fe/Mg$_{1-x}$Al$_{x}$O/Fe ($x=0.09$), and Fe/Mg$_{1-x}$Zn$_{x}$O/Fe ($x=0.025$), all with 6 atomic layers in the barrier; (c) and (d) $I$-$V$ curves for the same configurations as (a) and (b) but with antiparallel moments.}
\end{figure}

The calculated $I$-$V$ curves for the AlO MTJ exhibit strong NDR, as depicted in Fig. \ref{IV}(a) and (c), for both parallel and anti-parallel states.
For the MgAlO MTJ, there is also NDR for the parallel state, shown in Fig. \ref{IV}(b), but no NDR in the antiparallel state, shown in Fig. \ref{IV}(d).
For the MgZnO MTJ, the band bending is significant but the gap is much larger and no NDR is predicted. For AlO, NDR occurs in different bias regions for parallel and anti-parallel states, 0.8 V to 1.6 V for parallel and 1.4 V to 1.8 V for anti-parallel. For MgAlO, NDR only occurs for the parallel state.

Increasing the thickness of the barrier tends to suppress NDR, as shown in Fig. \ref{IV}(a) and (c). This results from the exponential decay of the MIGS charge in the barrier. Thus a large effect requires a thin barrier. This is in sharp contrast to the conventional approach of using a ferroelectric tunnel junction (FTJ) to change the band gap and tune its transport properties, where an
electroresistance is produced by switching the polarization direction of the ferroelectric layer \cite{Wen2013,Tsymbal2013,Garcia2009} and the effect is limited by a minimal critical thickness of the
layer \cite{Dawber2005}.

\subsection{Generalized Simmons formula for arched band gap}

\begin{figure}
\centering
\includegraphics[width=0.48\textwidth]{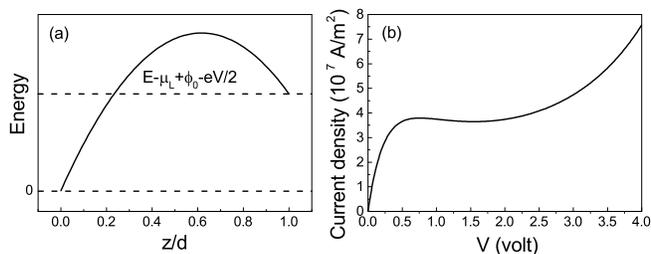}
\caption{\label{fig_simmons_model} Generalized two-band Simmons formula for an arched barrier, (a) effective barrier potential $U(z)$ in Eq. (3); (b) J-V curve for $a = 1$ eV nm$^{-2}$, $d = 1.5$ nm, and $\phi_0/e = 2$ volts. }
\end{figure}

Next we derive a generalized Simmons formula for a junction with an arched band gap. We invoke the WKB approximation for Eq. (\ref{eq2}),
\begin{equation}
T({\bf k}_{\|},E)=e^{-S\int^d_0\sqrt{eU(z)+\frac{\hbar^2k_{\|}^2}{2m}}dz},
\end{equation}
where $S=2\sqrt{2m}/\hbar = 10.25$ eV$^{-1/2}$nm$^{-1}$ and $eU(z)$ is assumed to have a simple form for an arched barrier,
\begin{equation}
eU(z)=az\left(d-z+\frac{E-\mu_L+\phi_0-eV/2}{ad}\right),
\end{equation}
with $\mu_L$ pinned at the conduction band edge at the left interface, $\phi_0$ being the effective barrier height measured from the valence band edge at the right interface under zero bias, and $a$ as a constant. A typical $U(z)$ is plotted in Fig. \ref{fig_simmons_model}(a). Defining $\phi_{\pm}=\phi_0\pm\frac{1}{2}eV $, we find,
\begin{equation}
J\approx \frac{16ead}{3\pi h S}\frac{e^{-\frac{S d}{4}\left(\sqrt{a}d+\sqrt{\phi_+}+\frac{\phi_+}{\sqrt{a}d}\right)}}{\sqrt{\phi_+}\left(d+\frac{\phi_+}{ad}\right)^2}
\left(e^{\frac{S}{4ad}\phi_+^{3/2}}-e^{\frac{S}{4ad}\phi_-^{3/2}}\right).
\label{Simmons}
\end{equation}
Fig. \ref{fig_simmons_model}(b) plots an example of $J(V)$ as a function of $V$, with the NDR appearing near $V\approx (1/2) \phi_0/e=1$ volt.

\section{Summary}

In summary, spin-dependent NDR is predicted in MTJs containing CCDC barrier layers of AlO$_x$ and Mg$_{1-x}$Al$_x$O by first principle calculation with MIGS model. Theoretical calculation also predicts the absence of NDR in Mg$_{1-x}$Zn$_x$O due to its larger band gap.
A general mechanism for producing NDR in ultrathin junctions without requiring magnetism is proposed for arched barriers.
The strong spin dependence of the NDR can be exploited for spin current amplification in devices or other spintronic devices, where when an ac signal is applied on top of a dc
bias set to the NDR voltage, the ac amplitude of the spin channel with NDR will gain by $r/(r-R)$ with $R$ being the circuit resistance and $-r$ being the slope of the NDR part of the I-V curve.
This work also provides a possible qualitative explanation for NDR observed in bilayer graphene \cite{Kim}. In this case a small gap of 0.14 eV is opened at the Dirac point ($E_D$) due to the built-in electric field from the SiC substrate.

\acknowledgments
This work was supported by the 863 Plan Project of Ministry of Science and Technology (MOST) (Grant No. 2014AA032904), the MOST National Key Scientific Instrument and Equipment Development Projects [Grant No. 2011YQ120053], the National Natural Science Foundation of China (NSFC) [Grant No. 11434014 \& 51229101] and the Strategic Priority Research Program (B) of the Chinese Academy of Sciences (CAS) [Grant No. XDB07030200].


\begin{thebibliography}{0}

\bibitem{dragoman2014negative}
  \Name{M. Dragoman, A. Dinescu \and D. Dragoman}
  \REVIEW{Nanotechnology}{25}{2014}{415201}

\bibitem{zhang2014negative}
  \Name{R. Shen, D. Wang, P. Tao, Y. Liu, X. Xia, Y. Luo, \and G. Du}
  \REVIEW{Appl. Phys. Lett.}{104}{2014}{053507}

\bibitem{sangwan2015gate}
  \Name{V. ~K. Sangwan, D.Jariwala, I. ~S. Kim, K. -S. Chen, T. ~J. Marks, L. ~J. Lauhon, \and M. ~C. Hersam}
  \REVIEW{Nature nanotechnology}{10}{2015}{403}

\bibitem{PhysRevB.83.115414}
  \Name{R. H\"artle \and M. Thoss}
  \REVIEW{Phys. Rev. B}{83}{2011}{115414}

\bibitem{xie2015negative}
  \Name{F. Xie, Z. -Q. Fan, K. Liu, H. -Y. Wang, J. -H. Yu, \and K. -Q. Chen}
  \REVIEW{Organic Electronics}{27}{2015}{41}

\bibitem{yang2014negative}
  \Name{S. Yang, P. Liu, S. Guo, L. Zhang, D. Yang, Y. Jiang, \and B. Zou}
  \REVIEW{Appl. Phys. Lett.}{104}{2014}{033301}

\bibitem{nguyen2015negative}
  \Name{P. ~D. Nguyen, T. ~C. Nguyen, F. ~M. Hossain, D. ~H. Huynh, R. Evans, \and E. Skafidas}
  \REVIEW{Nanoscale}{7}{2015}{289}

\bibitem{Esaki}
  \Name{L. Esaki}
  \REVIEW{Phys. Rev.}{109}{1958}{603}

\bibitem{Gunn1964}
  \Name{J.~B. Gunn}
  \REVIEW{{IBM} Journal of Research and Development}{8}{1964}{141}

\bibitem{esaki1966new}
  \Name{L. Esaki \and P. Stiles}
  \REVIEW{Phys. Rev. Lett.}{16}{1966}{1108}

\bibitem{Tsu1973}
  \Name{R. Tsu}
  \REVIEW{Appl. Phys. Lett.}{22}{1973}{562}

\bibitem{Chang1974}
  \Name{L. ~L. Chang}
  \REVIEW{Appl. Phys. Lett.}{24}{1974}{593}

\bibitem{BandgapGraphene}
  \Name{M. ~Y. Han, B. \"Ozyilmaz, Y. Zhang \and P. Kim}
  \REVIEW{Phys. Rev. Lett.}{98}{2007}{206805}

\bibitem{GrapheneTransistors}
  \Name{X. Wang,Y. Ouyang, X. Li, H. Wang, J. Guo \and H. Dai}
  \REVIEW{Phys. Rev. Lett.}{100}{2008}{206803}

\bibitem{GrapheneTransport}
  \Name{B. \"Ozyilmaz, P. Jarillo-Herrero, D. Efetov, D.~A. Abanin, L.~S. Levitov \and P. Kim}
  \REVIEW{Phys. Rev. Lett.}{99}{2007}{166804}

\bibitem{BiasedBilayerGraphene}
  \Name{E. ~V. Castro, et al.}
  \REVIEW{Phys. Rev. Lett.}{99}{2007}{216802}

\bibitem{Kim}
  \Name{K. ~S. Kim, T. -H. Kim, A. ~L. Walter, T. Seyller, H. ~W. Yeom, E. Rotenberg \and A. Bostwick}
  \REVIEW{Phys. Rev. Lett.}{110}{2013}{036804}

\bibitem{PhysRevB.73.033409}
  \Name{I. Weymann \and J. Barna\ifmmode~\acute{s}\else \'{s}\fi{} }
  \REVIEW{Phys. Rev. B}{73}{2006}{033409}

\bibitem{PhysRevLett.94.207210}
  \Name{Z. -Y. Lu, X. -G. Zhang, S. ~T. Pantelides}
  \REVIEW{Phys. Rev. Lett.}{94}{2005}{207210}

\bibitem{CiorgaAPL2002}
  \Name{M. Ciorga, M. Pioro-Ladriere, P. Zawadzki, P. Hawrylak \and A. ~S. Sachrajda}
  \REVIEW{Appl. Phys. Lett.}{80}{2002}{2177}

\bibitem{PhysRevB.32.6968}
  \Name{J. Tersoff}
  \REVIEW{Phys. Rev. B}{32}{1985}{6968}

\bibitem{PhysRevLett.114.146804}
  \Name{T. Iffl\"ander, S. Rolf-Pissarczyk, L. Winking, R. ~G. Ulbrich, A. Al-Zubi, S. Bl\"ugel \and M. Wenderoth}
  \REVIEW{Phys. Rev. Lett.}{114}{2015}{146804}

\bibitem{PhysRevB.63.054416}
  \Name{W. ~H. Butler, X. -G. Zhang, T. ~C. Schulthess \and J. ~M. MacLaren}
  \REVIEW{Phys. Rev. B}{63}{2001}{054416}

\bibitem{Parkin2004}
  \Name{S. ~S. ~P. Parkin, C. Kaiser, A. Panchula, P. ~M. Rice, B. Hughes, M. Samant \and S. -H. Yang}
  \REVIEW{Nat Mater}{3}{2004}{862}

\bibitem{Yuasa2004}
  \Name{S. Yuasa, T. Nagahama, A. Fukushima, Y. Suzuki and K. Ando}
  \REVIEW{Nat Mater}{3}{2004}{868}

\bibitem{Li2014srep}
  \Name{D. ~L. Li, et al.}
  \REVIEW{Scientific Reports}{4}{2014}{7277}

\bibitem{ZhangJiaAPL}
  \Name{J. Zhang, X. -G. Zhang \and X. ~F. Han}
  \REVIEW{Appl. Phys. Lett.}{100}{2012}{222401}

\bibitem{spinelAlO}
  \Name{Jensen Thomas N., et al.}
  \REVIEW{Phys. Rev. Lett.}{113}{2014}{106103}

\bibitem{PhysRevB.86.184401}
  \Name{H. Sukegawa, Y. Miura, S. Muramoto, S. Mitani, T. Niizeki, T. Ohkubo, K. Abe, M. Shirai, K. Inomata \and K. Hono}
  \REVIEW{Phys. Rev. B}{86}{2012}{184401}

\bibitem{simmons1963generalized}
  \Name{J. ~G. Simmons}
  \REVIEW{Journal of Applied Physics}{34}{1963}{1793}

\bibitem{PhysRevB.59.5470}
  \Name{J. ~M. MacLaren, X. -G. Zhang, W. ~H. Butler \and X. Wang}
  \REVIEW{Phys. Rev. B}{59}{1999}{5470}

\bibitem{vosko1980accurate}
  \Name{S. Vosko, L. Wilk \and M. Nusair}
  \REVIEW{Canadian Journal of physics}{58}{1980}{1200}

\bibitem{buttiker1985generalized}
  \Name{M. B{\"u}ttiker, Y. Imry, R. Landauer \and S. Pinhas}
  \REVIEW{Phys. Rev. B}{31}{1985}{6207}

\bibitem{PhysRevB.5.2382}
  \Name{B.~L. Gyorffy}
  \REVIEW{Phys. Rev. B}{5}{1972}{2382}

\bibitem{PhysRevB.69.134406}
  \Name{C. Zhang, X. -G. Zhang, P. ~S. Krsti\ifmmode~\acute{c}\else \'{c}\fi{}, H. -p. Cheng, W. ~H. Butler \and J.~M. MacLaren}
  \REVIEW{Phys. Rev. B}{69}{2004}{134406}

\bibitem{PhysRevB.80.144101}
  \Name{I. ~V. Maznichenko, et al.}
  \REVIEW{Phys. Rev. B}{80}{2009}{144101}




\bibitem{Wen2013}
  \Name{Z. Wen, C. Li, D. Wu, A. Li \and N. Ming}
  \REVIEW{Nat Mater}{12}{2013}{617}

\bibitem{Tsymbal2013}
  \Name{E. ~Y. Tsymbal \and A. Gruverman}
  \REVIEW{Nat Mater}{12}{2013}{602}

\bibitem{Garcia2009}
  \Name{V. Garcia, S. Fusil, K. Bouzehouane, S. Enouz-Vedrenne, N. ~D. Mathur, A. Barthelemy \and M. Bibes}
  \REVIEW{Nature}{460}{2009}{81}

\bibitem{Dawber2005}
  \Name{M. Dawber, K. ~M. Rabe \and J. ~F. Scott}
  \REVIEW{Reviews of Modern Physics}{77}{2005}{1083}

\end{thebibliography}
\end{document}